\documentclass{ptapap}

\author{Oleg Kochukhov}[UU]
\affil[UU]{Department of Physics and Astronomy, Uppsala University, Box 516, SE-75120 Uppsala, Sweden}

\title{Mapping Stellar Magnetic Fields}

\begin{document}

\maketitle

\begin{abstract}

This review discusses the problem of reconstruction of surface magnetic field topologies of early-type stars with a focus on mapping methods utilising information content of high-resolution spectropolarimetric observations. Basic principles of the Zeeman Doppler imaging tomographic mapping technique are outlined and its recent applications to magnetic early-type stars are summarised. The current observational and modelling challenges faced by the studies of surface magnetic fields in these stars are also discussed.

\end{abstract}

\section{Introduction}

Magnetic field is an important agent affecting formation and evolution of stars. The presence of a magnetic field in stars is often difficult to recognise, yet its effects may be profound and even crucial for certain evolutionary phases. There are two basic types of stellar magnetic field and associated magnetic activity phenomenology. Late-type stars with convective outer envelopes (spectral types FGKM on the main sequence) generate magnetic fields through a contemporary dynamo process \citep{charbonneau:2013} -- a conversion of the rotational and turbulent kinetic energy of plasma flows into magnetic field energy. This process gives rise to complex, relatively weak ($\sim$\,1--100 G disk-averaged field strength) surface magnetic field structures. These fields evolve on time scales ranging from weeks (emergence and decay of active regions) to decades (activity cycles). Magnetism of cool stars powers a multitude of high-energy, often transient, variability phenomena (flares, spot-induced photometric rotational modulation, non-thermal chromospheric and coronal emission) and is tightly correlated with stellar rotation. Young rapidly rotating stars display stronger magnetic fields and exhibit correspondingly higher levels of magnetic activity. As cool stars age and shed their angular momentum, their dynamos become less efficient, leading to weaker surface magnetic fields and less prominent activity phenomena. All single late-type stars follow this evolutionary path. Therefore, all of them can be deemed magnetic to some extent.

In contrast, stars with radiative envelopes (spectral types OBA on the main sequence) exhibit an entirely different magnetic activity phenomenology. Only a small fraction, on the order of 10\%, of these stars are observably magnetic \citep{alecian:2013,wade:2016}. The incidence of magnetism does not correlate with particular evolutionary stages and does not depend on stellar rotation. Magnetic early-type stars possess relatively strong (disk-averaged field strengths from a few hundred G to $\sim$\,30~kG), globally-organised magnetic fields. These fields shape non-uniform distributions of chemical elements in BA stars \citep{michaud:1981,alecian:2015} and circumstellar material in early-B and O stars \citep{townsend:2005a}. A remarkable feature of these magnetic topologies and associated surface and circumstellar structures is their long-term stability. No changes of the magnetic field geometries or chemical spot morphologies have been conclusively established for any early-type stars, making contemporary dynamo an unlikely source of magnetism. Instead, it is believed that these fields are stable remnants \citep{braithwaite:2006} of the magnetic flux generated or acquired by stars at earlier evolutionary phases. Different processes could be responsible for the origin of these fossil magnetic fields, for example dynamo operating in the short-lived convective part of pre-main sequence evolution \citep{moss:2001,shultz:2019} or binary star mergers \citep{schneider:2019}. None of these hypothesis succeeded in explaining nearly constant incidence rates, typical strengths and topological properties of fossil magnetic fields across the entire range of masses and evolutionary stages where these fields are detected.

Reliable constraints on the strengths and detailed information on the geometries of stellar surface magnetic fields is a prerequisite for developing stellar magnetism theories and understanding complex relationships between magnetic fields, stellar variability and surface structure formation. This puts the problem of stellar surface mapping in the centre of ongoing research on magnetically active cool and hot stars. In this review I provide a historical perspective on the development of methods of mapping magnetic fields in early-type stars. I also summarise results of applications of modern variants of these methods to high-resolution spectropolarimetric observations and discuss reliability of these results.

\section{A brief history of stellar magnetic field mapping}

Development of the oblique rotator framework \citep{stibbs:1950} was the starting point of the research aimed at understanding the structure of stellar magnetic fields and its connection to inhomogeneous chemical element distributions. In the initial version of this phenomenological model magnetic field geometry was described with an oblique poloidal dipolar field frozen within the outer stellar envelope. Related surface and circumstellar inhomogeneities were treated with various levels of geometrical detail, but were invariably assumed to  be stationary and co-rotating with the star and its magnetic field. This basic framework of interpretation of observations of magnetic early-type stars survives to this day.

In early applications of the oblique rotator model parameters of dipolar field were adjusted by fitting phase curves of the mean longitudinal magnetic field \citep{borra:1980} and, less commonly, mean field modulus \citep{huchra:1972,borra:1978}. Theoretical curves of magnetic observables were produced with a straightforward geometrical integration of the surface vector field corresponding to a given combination of dipolar parameters. Occasionally, interpretation of high-quality observations called for a more complex field topologies such as offset dipoles \citep{borra:1977} or superposition of dipolar and quadrupolar components \citep{landstreet:1990}. This methodology evolved into multipolar fitting of magnetic observables (see below), which to this day remains the main source of basic information on the magnetic field properties of large samples of early-type stars.

It was recognised early on that high-resolution measurements of circular and linear polarisation profiles of spectral lines provide an ultimate source of information about stellar magnetic fields \citep{borra:1973,borra:1977a}, potentially ushering more detailed and reliable field geometry models. Attempts to reproduce such high-quality polarisation observations with theoretical calculations were limited to a few brightest stars \citep{borra:1980b,glagolevskii:1985a}. Nevertheless, these studies succeeded in confirming the general validity of the oblique rotator model and allowed to unify magnetic field and chemical spot modelling into a single stellar surface mapping procedure based on numerical solution of a regularised inverse problem.

In the next important step, Landstreet and collaborators \citep{landstreet:1988,landstreet:1989} pioneered methods for calculating stellar four Stokes parameter spectra by solving polarised radiative transfer equation in realistic model atmospheres. This effort yielded the {\tt Zeeman} code \citep{wade:2001}, which provided computational framework for detailed interpretation of time-resolved, high-resolution observations of line profiles strongly distorted by magnetic field. This modelling approach, combined with fitting phase variation of the longitudinal field and mean field modulus, allowed to deduce both magnetic field topology (represented by a superposition of co-aligned dipole, quadrupole, and octupole) and a schematic distribution of chemical elements (assumed to be axisymmetric with respect to the magnetic field axis). Calculations with {\tt Zeeman} are still frequently used today to account for the effects of chemical inhomogeneities and magnetic field on the intensity spectra of early-type stars \citep{bailey:2011,bailey:2012}, and, less commonly, to interpret high-resolution polarisation observations \citep{khalack:2006}.

New generation of night-time high-resolution spectropolarimeters enabled studies of four Stokes parameter spectra for a large number of early-type magnetic stars \citep{donati:1997,wade:2000b,silvester:2012}. Taking advantage of these observational data, magnetic (Zeeman) Doppler imaging technique was independently developed for both late-type \citep{brown:1991,hussain:2000,kochukhov:2013} and early-type \citep{piskunov:2002a} stars. Specific implementations of the MDI/ZDI inversion techniques differ in many essential details, such as degree of sophistication of calculation of theoretical local Stokes parameter profiles, self-consistency between magnetic and star spot mapping, and types of observational data (individual lines or average profiles) used as input. However, a common feature of all these procedures is that they reconstruct stellar magnetic topology directly from polarisation profiles and use many degrees of freedom to describe arbitrary complex magnetic geometries, without restricting them to low-order, poloidal multipoles.

\section{Multipolar fits to integral magnetic observables}

In its most basic form multipolar fitting can be employed to determine dipolar field strength $B_{\rm d}$ and magnetic obliquity angle $\beta$ from variation of longitudinal magnetic field $\langle B_{\rm z} \rangle$. This magnetic observable provides a measure of the disk-integrated, line of sight magnetic field, weighted by the continuum and line intensity. Fitting $\langle B_{\rm z} \rangle$ curves has a major advantage in its ability to incorporate any type of longitudinal field measurement, thus benefiting from large volumes of historic data \citep{bychkov:2005} as well as observational material supplied by modern low- \citep{bagnulo:2015} and high-resolution \citep{morel:2015,wade:2016} spectropolarimetric surveys. Dipolar fits of $\langle B_{\rm z} \rangle$ data enable analysis of statistically unbiased stellar samples \citep{auriere:2007,sikora:2019a} and currently represent the only practical approach to the problem of studying evolutionary changes of magnetic field characteristics of early-type stars \citep{shultz:2019a}.

Studies based on dipolar modelling of $\langle B_{\rm z} \rangle$ curves have established that polar field strengths of B and A stars follow a log-normal distribution with the most probable $B_{\rm d}$ of about 2.5~kG and revealed a lower cutoff at $B_{\rm d}\approx300$~G \citep{auriere:2007,sikora:2019a}. There is an evidence of a long-term decay of the dipolar field component. On the other hand, magnetic field axes appear to be randomly oriented \citep{sikora:2019a} and does not correlate with stellar parameters or other field characteristics.

Various generalisations of dipolar fitting to more complex field configurations have been proposed \citep[e.g.][]{hensberge:1977,khokhlova:2000,khalack:2003}, however only two such approaches were systematically applied to real observations of many magnetic stars \citep{landstreet:2000,bagnulo:2002}. Both multipolar field mapping methods combine interpretation of $\langle B_{\rm z} \rangle$ and field modulus variation with fitting several line profile moments (integral magnetic observables) obtained from medium-resolution circular polarisation observations with the so-called moment technique \citep{mathys:1986}. \citet{landstreet:2000} reproduced observations of longitudinal field, field modulus, mean quadratic field and crossover \citep{mathys:1995,mathys:1995a} with a superposition of co-aligned dipole, quadrupole and octupole. \citet{bagnulo:2002} fitted the same data with a non-axisymmetric combination of dipolar and non-linear quadrupolar fields. Both studies found evidence of a systematic change of the dipolar field obliquity with the stellar rotational period, with low-$\beta$ geometries being more common in slow rotators. On the other hand, the authors reached different conclusions regarding the dominant field component despite considering essentially the same observational data. \citet{landstreet:2000} suggested that the dipolar contribution is typically stronger than the quadrupolar one in most of the stars. In contrast, \citet{bagnulo:2002} found dominant quadrupolar fields for the majority of the targets.

Notwithstanding many useful results obtained with the multipolar fitting technique and the ease of its practical application to large stellar samples, this method is very restrictive and biased in several significant aspects. For example, the choice of specific low-order poloidal field parameterisation and complete neglect of toroidal fields is subjective and not physically motivated. This choice also represents a major source of non-uniqueness of the final results since application of different parameterisations to the same data sets often leads to entirely different magnetic field geometries \citep{kochukhov:2004d,kochukhov:2006c}. Furthermore, this method cannot easily incorporate effects of non-uniform horizontal distributions of chemical elements, which in many cases lead to major changes of the shape and amplitude of $\langle B_{\rm z} \rangle$ curves depending on chemical element \citep[e.g.][]{rusomarov:2013,yakunin:2015,shultz:2018a}. Thus, there is always an ambiguity in interpretation of the deviations of observations from best-fitting model curves: are these deviations appear due to chemical spots or due to departures of the field structure from the assumed low-order multipolar configuration? Finally, a successful fit of first few line profile moments by no means guarantees that the original Stokes parameter profiles are also well reproduced. Several studies found that polarisation profiles corresponding to seemingly satisfactory multipolar fits of integral magnetic observables fail to provide a reasonable match to the observed Stokes parameter spectra \citep{bagnulo:2001,kochukhov:2011,kochukhov:2017a}. The degree of this problem varies from moderate to severe, depending on the star. This indicates that certain aspects of stellar magnetic field geometries cannot be captured, even in principle, by the classical integral magnetic observables.

\section{Zeeman Doppler imaging with Stokes parameter spectra}

\subsection{Methodology}

Doppler and Zeeman Doppler imaging (DI and ZDI) are indirect stellar surface mapping techniques that extract information on stellar surface structures from observation of rotational modulation of spectral line profiles \citep{kochukhov:2016}. These tomographic techniques are based on the fact that a structure on the stellar surface, for example a cool spot, produces local spectral profile distinct from the one corresponding to the surrounding photosphere. If this spot is sufficiently large, its spectral contribution is observed as a distortion in the disk-integrated line profile, propagating from blue to red in the course of stellar rotation. Doppler imaging converts a set of time-resolved line profiles, representing instantaneous one-dimensional projections of the stellar surface, into a two-dimensional map. Depending on how this map is parameterised, DI requires an additional mathematical constraint -- regularisation -- to ensure uniqueness of the solution.

Both DI and ZDI perform best for fast rotating stars since in that situation they benefit from longitudinal resolution of the stellar surface by rotational Doppler broadening. In the limit of slow rotation, when the local line width is comparable or larger than the projected rotational velocity, DI becomes less useful as it essentially degenerates into fitting an equivalent width phase curve. On the other hand, ZDI can still meaningfully constrain magnetic field topology of slow rotators drawing information from significant rotational modulation of amplitudes and shapes of polarisation signatures \citep{kochukhov:2002c,kochukhov:2016a}. In fact, ZDI has been successfully applied to many sharp-lined solar-type stars \citep{petit:2008a,morgenthaler:2012,rosen:2016,jeffers:2017} yielding, for example, the first evidence of magnetic polarity reversals associated with stellar activity cycles in stars other than the Sun \citep{boro-saikia:2018}.

An important feature of any implementation of ZDI, which sets it aside from the multipolar fitting methods discussed above, is availability of many more degrees of freedom to describe stellar magnetic field topologies. In the early ZDI studies of magnetic Ap stars \citep{kochukhov:2002b,kochukhov:2004d,kochukhov:2011a} the field structure was represented by three independent rectangular maps of the magnetic field vector components. Later investigations \citep{kochukhov:2014,kochukhov:2017a,kochukhov:2019,rusomarov:2016,rusomarov:2018,silvester:2015,silvester:2017} switched to using a general spherical harmonic parameterisation, which treats a surface vector field distribution as a superposition of poloidal and toroidal harmonic components with arbitrary degree of complexity. The latter is controlled by setting the maximum angular degree $\ell$ of the spherical harmonic expansion and also using a harmonic penalty function \citep{morin:2008,kochukhov:2014} to prevent spurious contribution of high-order modes. High-quality spectropolarimetric observations of sharp-lined stars ($v_{\rm e}\sin i \le 5$~km\,s$^{-1}$) probe harmonic modes of up to $\ell_{\rm max}\approx5$. ZDI modelling of rapid rotators ($v_{\rm e}\sin i \ge 30$~km\,s$^{-1}$) justifies using $\ell_{\rm max}=20$--30. Harmonic parameterisation enables ZDI to address new scientific problems, such as quantifying relative contributions of toroidal and poloidal magnetic fields and studying small-scale magnetic structures that cannot be detected with low-order multipolar fits of line profile moments. At the same time, harmonic parameterisation ensures that reconstructed vector field maps are physically meaningful as they automatically satisfy Maxwell's equations.

ZDI codes differ in terms of the level of detail of local line profile modelling and the type of observational data they are applied to. The original ZDI/MDI methodology introduced for Ap stars by \citet{piskunov:2002a} relied on modelling individual spectral lines in Stokes $IV$ or all four Stokes parameters. Magnetic field and horizontal chemical inhomogeneities were treated self-consistently using single average stellar model atmosphere. A coupling between local abundance and atmospheric structure was explored in later studies \citep{kochukhov:2012,oksala:2015}, but was proven to be unnecessary except for DI analyses of He-rich stars. On the other hand, reliance on individual line profiles represents a significant limitation of practical ZDI application since it requires spectropolarimetric observations of a very high quality. Such data can currently be obtained for a small fraction of early-type magnetic stars. Lower quality spectra of more numerous fainter stars still reveal clear polarisation signatures in the average Stokes parameter profiles calculated with the least-squares deconvolution (LSD, \citealt{kochukhov:2010a}) method. It was, therefore, essential to extend ZDI to this type of observational data. One way to accomplish this is to assume that LSD Stokes $V$ profile behaves as a fictitious line with average parameters and a triplet Zeeman splitting \citep{donati:2006b}. Another approach, suggested by \citet{kochukhov:2014}, is to produce local LSD profile tables by averaging over theoretical Stokes spectra of large number of individual lines calculated with realistic model atmospheres and detailed polarised radiative transfer methods. The latter approach is the only viable option for reconstructing maps of chemical abundance spots and modelling linear polarisation (Stokes $Q$ and $U$) LSD profiles \citep{kochukhov:2010a}. This version of the ZDI technique, currently representing the state of the art in hot star magnetic mapping, was successfully applied in several recent studies of magnetic A and B-type stars \citep{kochukhov:2017a,kochukhov:2019,oksala:2018}.

\begin{figure}
\centering
\includegraphics[width=8.5cm, angle=270]{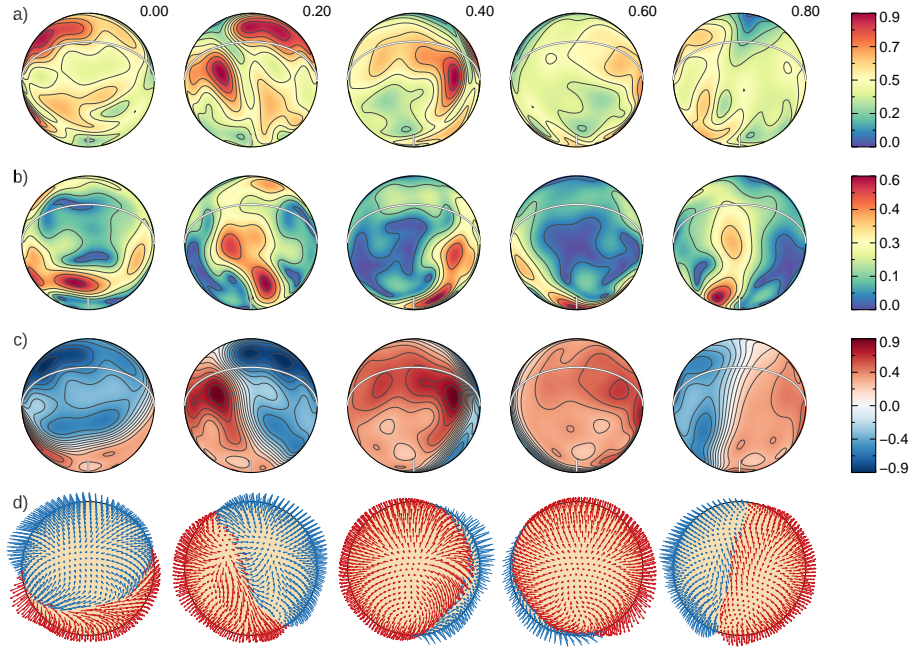}
\caption{Typical example of distorted dipolar magnetic field geometry obtained by ZDI study of a magnetic Ap star. These maps show magnetic field topology of $\theta$~Aur derived from Stokes $IQUV$ LSD profiles \citep{kochukhov:2019}. The star is shown at five rotation phases, as indicated above each column. The four rows of spherical plots show distributions of {\bf a)} field modulus, {\bf b)} horizontal field, {\bf c)} radial field, and {\bf d)} field orientation. The contours over spherical maps are plotted with a step of 0.1~kG. The colour bars give the field strength in kG. 
}
\label{fig1}
\end{figure}

\begin{figure}
\centering
\includegraphics[width=9.0cm]{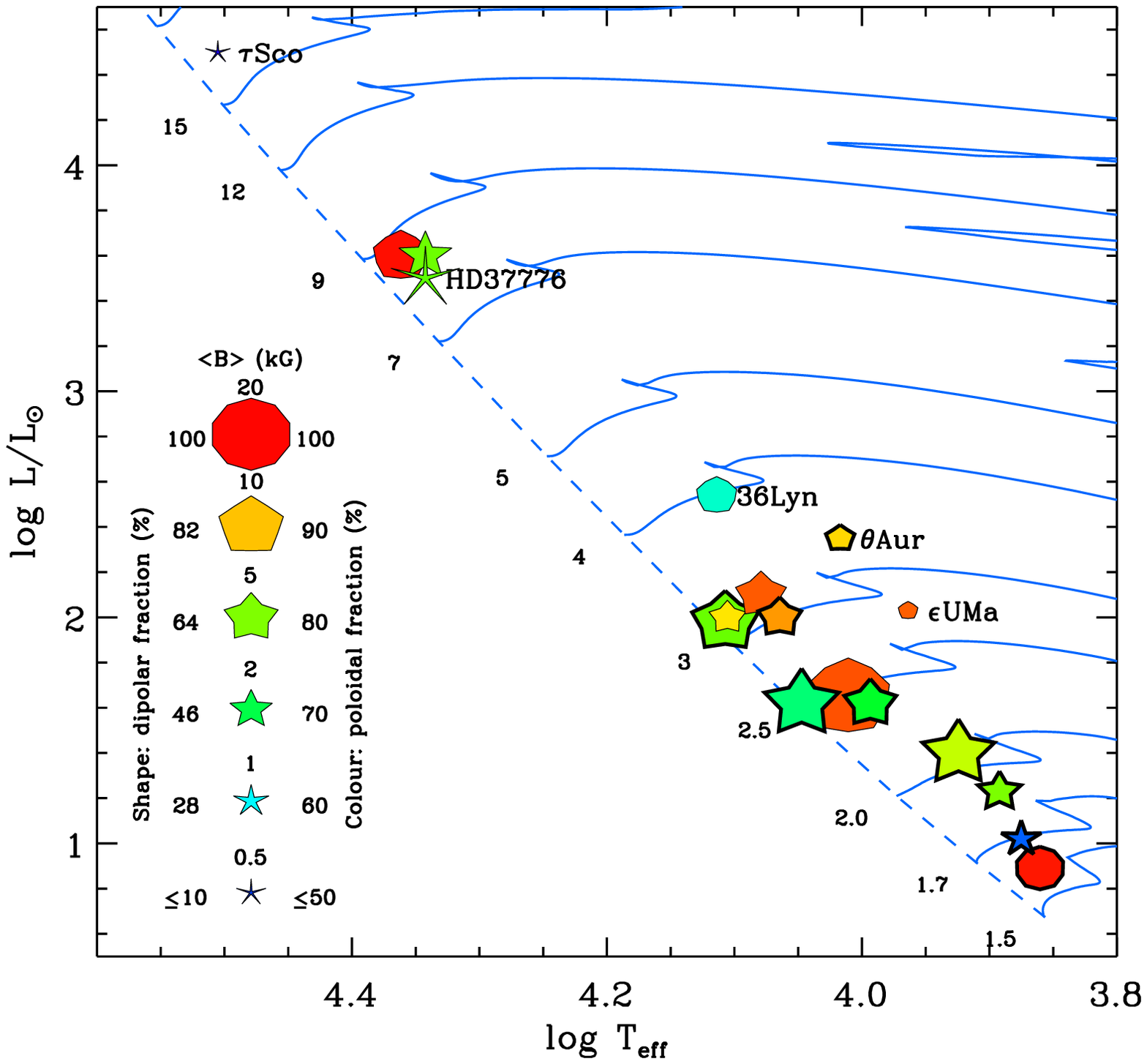}
\caption{Characteristics of global magnetic field topologies of early-type stars studied with ZDI. The symbol size indicates the average field strength. The symbol shape corresponds to  contribution of the dipolar component to the total magnetic field energy (from decagons for purely dipolar fields to pointed stars for non-dipolar field topologies). The symbol colour reflects contribution of the toroidal magnetic field component (red for purely poloidal geometries, dark blue for field configurations with $\ge$50\% toroidal field contribution). The thickness of the symbol outline indicates stars studied with full Stokes vector ZDI (thick outline) or using Stokes $IV$ inversions (thin outline). 
}
\label{fig2}
\end{figure}

\subsection{Recent results}

Application of ZDI to polarisation spectra of early-type stars led to a number of important discoveries. The presence of small-scale deviations from canonical oblique dipolar topologies was firmly established in some magnetic stars \citep{kochukhov:2010,silvester:2014,rusomarov:2018}, but not all of them \citep{rusomarov:2015}. When available, linear polarisation spectra typically suggested a larger deviation from dipolar fields than can be inferred from circular polarisation data alone \citep{kochukhov:2010,rusomarov:2018}. However, this gain of spatial resolution brought by (still very challenging and time-consuming) full Stokes vector magnetometry is mostly noticeable for slow rotators ($v_{\rm e}\sin i \le 20$~km\,s$^{-1}$), but vanishes for stars with $v_{\rm e}\sin i$ exceeding $\sim$\,30~km\,s$^{-1}$ \citep{kochukhov:2019}.

Interestingly, when it comes to large-scale deviations from dipolar topologies, ZDI studies concluded that surface magnetic fields of early-type stars are \textit{less} complex than  believed previously. Only two early-type stars, HD\,37776 \citep{kochukhov:2011a} and $\tau$~Sco \citep{donati:2006b,kochukhov:2016a}, with clearly non-dipolar, very complex magnetic field structures have been found. Analysis of several stars suspected of hosting predominantly quadrupolar fields \citep{silvester:2015,kochukhov:2017a} indicated that their magnetic topologies are better described by distorted dipoles. These results support the basic oblique rotator model (centred dipolar field) but question the most popular extensions of this model (superposition of dipole and quadrupole) used by multipolar fitting studies. In general, the most common magnetic field configuration of an early-type star seems to be a dipole distorted at both large (an overall offset or a contribution of toroidal field) and small (magnetic spots) scales. Figure~\ref{fig1} shows an example, taken from \citet{kochukhov:2019}, of typical stellar magnetic field geometry of this kind.

Another interesting finding of ZDI investigations is evidence of non-negligible and, in some cases, very large contributions of toroidal magnetic field at the stellar surface \citep{kochukhov:2016a,oksala:2018,rusomarov:2018}. Possibility of such magnetic field structures was disregarded by all previous low-order multipolar modelling studies. However, strong toroidal fields must exist in stellar interiors to ensure long-term dynamical stability of global fossil field geometries \citep{braithwaite:2006}. Thus, their presence on stellar surfaces is theoretically justified despite systematic neglect by previous observational studies.

Figure~\ref{fig2} summarises results of 18 modern ZDI studies of early-type stars. Several properties of global magnetic field geometries (average field strength, poloidal-to-toroidal component ratio, contribution of higher order harmonic components) are displayed in the temperature-luminosity plane. No clear trends emerge except, possibly, weaker fields in older stars. But dating field stars from their position in the Hertzsprung-Russell diagram is notoriously problematic \citep{landstreet:2007}, which may explain absence of any dependence of the field complexity on age. The latter could be explored much more reliably by carrying out ZDI analysis of magnetic stars in open clusters. There is, however, some evidence that the field complexity depends on mass since so far we found examples of decisively non-dipolar fields only in the most massive magnetic stars.

ZDI modelling of Stokes parameter spectra of Ap stars provided a new generation of chemical abundance maps reconstructed fully accounting for modifications of spectral lines by the Zeeman effect \citep{kochukhov:2017}. These accurate multi-element star spot distributions have not, however, led to a better concordance with theoretical atomic diffusion predictions \citep{leblanc:2009,alecian:2010,alecian:2015,stift:2016}. It turns out that the observed chemical abundance distributions are far more diverse and complex than ubiquitous accumulation of elements in the horizontal field regions predicted by the theory \citep{kochukhov:2018a}. These results suggest that magnetic field is not the only and, often, not even the main process shaping surface element inhomogeneities. ZDI abundance mapping results call for development of numerically more sophisticated and physically more realistic atomic diffusion models that include inhomogeneous mass loss, stellar rotation, and meridional circulation. Transient and time-dependent hydrodynamical effects altering atomic diffusion process should not be ignored either. Manifestation of these time-dependent effects has been probably observed as evolving low-contrast chemical spots on non-magnetic HgMn stars \citep{kochukhov:2007b,korhonen:2013}.

\subsection{Limitations and uncertainties}

It is essential to assess precision, reliability, and intrinsic limitations of any involved spectral inversion procedure. Several studies addressed these issues for previous generation of ZDI codes, which described magnetic fields with two-dimensional pixel maps \citep{donati:1997a,kochukhov:2002c}. While these investigations offered a number of useful insights, their conclusions cannot be directly extrapolated to modern ZDI analyses based on spherical harmonic treatment of magnetic geometries. 

Internal consistency of inversion results can be tested by comparing independent magnetic maps of the same star obtained from LSD profiles of different chemical elements. Using this method, \citet{kochukhov:2019} concluded that typical local precision of the recovery of radial, meridional, and azimuthal field vector components is $\approx$\,10\% of the maximum surface field strength. Local field inclination (a key parameter in modelling atomic diffusion processes in magnetic field) is consistent to within $\approx$\,10$^{\rm o}$. The global field characteristics, such as the fractions of poloidal vs. toroidal and dipolar vs. non-dipolar magnetic field energies, agree to within 2--8\% in independent maps. This study also did not find a significant degradation of the field mapping results for stars with $v_{\rm e}\sin i$\,=\,35--54~km\,s$^{-1}$ when Stokes $QU$ profiles were neglected in the magnetic inversion.

On the other hand, \citet{kochukhov:2016a} demonstrated that a combination of several compounding factors (slow stellar rotation, absence of linear polarisation data, unusually high degree of field complexity) can render a ZDI inversion highly ambiguous. In this study drastically different magnetic field maps were obtained for $\tau$~Sco by using slightly different versions of high-order harmonic field expansion. For this star linear polarisation observations appear to be essential for validating results inferred from the Stokes $V$ profile modelling.

\section{Conclusions and outlook}

Zeeman Doppler imaging is now a well-established technique of utilising full information content of high-resolution stellar polarisation spectra. It is the ultimate magnetic field mapping method, capable of delivering detailed and reliable magnetic field vector topologies as well as distributions of associated chemical inhomogeneities. However, current implementations of ZDI are extremely demanding in terms of computational effort and the quality of observational data. They also require an in-depth understanding of the absorption spectra and atmospheres of individual stars. These requirements hinder application of ZDI to large samples of early-type magnetic stars and make it impossible to model fainter objects for which polarisation is detected only in the LSD profiles generated from all metal lines. These methodological problems can be overcome by developing a modified ZDI technique that makes use of mean LSD profiles and employs simplified polarised radiative transfer schemes in place of the full numerical solution. It is likely that in the near future we will see a rise in applications of such simplified ZDI methodologies. This will help to map stellar magnetic topologies across the entire upper part of the Hertzsprung-Russell diagram occupied by intermediate-mass and massive stars, thereby establishing how the field topologies change with stellar mass and age.

\acknowledgements{I acknowledge support by the Knut and Alice Wallenberg Foundation, the Swedish Research Council, and the Swedish National Space Board.}

\bibliographystyle{ptapap}
\bibliography{astro_papers}

\end{document}